\documentclass[journal]{IEEEtran}
\usepackage{blindtext}
\usepackage{verbatim}
\usepackage{csquotes}

\usepackage{cite}

\usepackage{graphicx}
\graphicspath{ {media/} }

\usepackage{amsmath}
\usepackage{bm}
\newcommand{\BigO}[1]{\ensuremath{\operatorname{O}\bigl(#1\bigr)}}

\usepackage{amssymb}
\usepackage{array}

\begin{document}
%
\title{WLS-Based Self-Localization Using Perturbed Anchor Positions and RSSI Measurements}

\author{Vikram~Kumar,~Reza~Arablouei,~Brano~Kusy,~Raja~Jurdak,~and~Neil~W.~Bergmann%

\thanks{V. Kumar and N. W. Bergmann are with the School of Information Technology and Electrical Engineering, University of Queensland, St Lucia QLD 4067, Australia (email: v.kumar@uq.edu.au; neil.bergmann@itee.uq.edu.au).}%

\thanks{R. Arablouei, B. Kusy, and R. Jurdak are with the Commonwealth Scientific and Industrial Research Organisation, Pullenvale QLD 4069, Australia (email: \{reza.arablouei, brano.kusy, raja.jurdak\}@csiro.au).}
}

\maketitle 
\begin{abstract}
We consider the problem of self-localization by a resource-constrained node within a network given radio signal strength indicator (RSSI) measurements from a set of anchor nodes where the RSSI measurements as well as the anchor position information are subject to perturbation. In order to achieve a computationally efficient estimate for the unknown position, we minimize a weighted sum-square-distance-error cost function in an iterative fashion utilizing the gradient-descent method. We calculate the weights in the cost function by taking into account perturbations in both RSSI measurements and anchor node position information while assuming normal distribution for the perturbations in the anchor node position information and log-normal distribution for the RSSI-induced distance estimates. The latter assumption is due to considering the log-distance path-loss model with normally-distributed perturbations for the RSSI measurements in the logarithmic scale. We also derive the Cramer-Rao lower bound associated with the considered position estimation problem. We evaluate the performance of the proposed algorithm considering various arbitrary network topologies and compare it with an existing algorithm that is based on a similar approach but only accounts for perturbations in the RSSI measurements. The experimental results show that the proposed algorithm yields significant improvement in localization performance over the existing algorithm while maintaining its computational efficiency. This makes the proposed algorithm suitable for real-world applications where the information available about the positions of anchor nodes often suffer from uncertainty due to observational noise or error and the computational and energy resources of mobile nodes are limited, prohibiting the use of more sophisticated techniques such as those based on semidefinite or second-order cone programming.

\end{abstract}


\IEEEpeerreviewmaketitle

\section{Introduction}
Location information is important in most sensor network applications such as environmental, wildlife, or mobile asset monitoring. The global positioning system (GPS) has revolutionized location-based services in the past few decades offering an accuracy range of 1-50m with consumer-grade GPS receivers in outdoor environments \cite{Vallina-Rodriguez2013}. However, the GPS is a relatively power-hungry technology and localization using the GPS is not practical in resource-constrained scenarios such as those involving wireless sensors networks (WSNs) with battery-operated mobile sensors \cite{Abdesslem2009,constandache2009enloc}. Relatively high cost and size of the currently available GPS receivers is another hindrance to this technology in becoming the primary choice of localization in most resource-constrained applications. 

Cooperative localization based on radio signal strength indicator (RSSI) measurements is an effective method for localizing groups of battery-operated mobile sensors  \cite{Jurdak2012,lee2012comon,mohammadabadi2014cooperative}. In cooperative localization, any node interested in estimating its own position, referred to as the \enquote{blind node}, receives position and distance information from its neighboring nodes, called the \enquote{anchor nodes}, and estimates its position through multilateration. Such localization techniques can be energy/cost efficient by minimizing the use of expensive localization methods such as the GPS. The localization performance of these techniques largely depends on the accuracy of the RSSI measurements and the anchor node position information. In practice, the RSSI measurements are subject to perturbations arising from model inaccuracy, thermal noise, measurement error, etc. The anchor node position information is also corrupted by noise or error, especially when it is the product of a previous estimation process including the GPS.

Substantial research effort has been dedicated to develop localization methods that can compensate for the adverse effects of perturbations in the RSSI measurements \cite{Feng2010, patwari2003relative, patwari2005locating, Tarrio2011, Shang2004, Biswas2004} and the anchor node position information \cite{zheng2009robust, Picard2010, Zhu2009, lui2009semi, Mekonnen2014, Naddafzadeh-Shirazi2014, srirangarajan2007distributed, Angjelichinoski2014, Costa2006, Ho2004}. Most of the proposed methods that can cope with perturbations in both RSSI measurements and anchor node position information rely on optimization techniques such as semidefinite programming (SDP) or second-order cone programming (SOCP). Therefore, high computational complexity of the SDP and SOCP often makes them unsuitable for localization in resource-constrained applications.

In this paper, we propose a computationally efficient algorithm to localize a blind node when perturbations are present in both RSSI measurements and anchor node position information. Our algorithm is of weighted least-squares (WLS) type as it is based on minimizing a weighted sum-square-error cost function using the gradient-descent method. Each term in the cost function is the weighted square of the difference between the distance inferred from an RSSI measurement and the Euclidean distance between the blind node and the anchor node to which the RSSI measurement corresponds. We weight the square-error terms by their variances, which are estimated by taking into account perturbations in both RSSI measurements and anchor node position information. We assume normal distribution for the perturbations in the available information of the anchor node positions. In addition, we adopt the well-known log-normal shadowing path-loss model. Therefore, we assume that the RSSI measurements in the logarithmic (dBm) scale are affected by normally-distributed perturbation and consequently the distance estimates inferred from the RSSI measurements have log-normal distribution. We derive the Cramer-Rao lower bound (CRLB) on the variance of any unbiased estimator of the blind node position in the considered problem with the assumed realistic distributions for the perturbations. We carry out numerical simulations with various geometrical arrangements of the blind node and the anchor nodes in arbitrary network topologies to evaluate the performance of the proposed algorithm in comparison with an existing WLS-based algorithm \cite{Tarrio2011} that only takes into account the perturbations in the RSSI measurements and has similar computational requirements. The simulation results show that the proposed algorithm offers significant improvement in the localization performance over the algorithm of \cite{Tarrio2011}.

Our work is motivated by several use-case scenarios relevant to localization in emerging cooperative wireless networks of mobile nodes such as robots, vehicles, and equipment or wildlife trackers deployed in large outdoor areas \cite{Jurdak2012, Atts2015, Allen2015, Juang2002, ahmad2015experiments}. Typically, such networks are very large and contain a mixture of GPS-equipped and non-GPS-equipped nodes. GPS-equipped nodes generally act as anchor nodes for the rest of the network. However, poor performance of the GPS in dense forests, urban areas, or even with a cloudy sky results in inaccuracies/uncertainties in the form of perturbations in the anchor node position information.

The remainder of the paper is organized as follows. After describing the notations used in this paper, in Section II, we review the existing works on the problem of self-localization in WSNs and highlight their differences with our work. For the sake of completeness, we also cover some works on the closely-related problem of source/emitter localization in WSNs. In Sections III and IV, we provide a formal statement of the considered problem and the details of the proposed algorithm, respectively. In Section V, we calculate the CRLB. We explain the experimental setup used for performance evaluation of the proposed algorithm in Section VI and provide the results of the experiments in Section VII before concluding the paper in Section VIII. 

\textit{Notations}: The symbol $\mathbb{N}^+$ denotes the set of positive integers and $\mathbb{R}$ denotes the set of real numbers. A lower-case letter, e.g., $x$, represents a scalar variable; a lower-case bold letter, e.g., $\mathbf{r}$, represents a vector; and an upper-case bold letter, e.g., $\mathbf{F}$, represents a matrix. The superscript $(.)^T$ denotes the vector/matrix transpose and ${(.)^{-1}}$ denotes the matrix inverse. A letter with a tilde accent, e.g., $\tilde{x}_{i}$, represents the noisy observation of the original variable, $ {x}_{i}$; a letter with a hat accent, e.g., $\hat{x}$, represents an estimated value; and a letter with an overbar, e.g., $\bar{e}$, represents an approximate value. The operator $\frac{\partial(.)}{\partial v}$ is the partial derivative with respect to the variable $v$, $\mathrm{Var}(.)$ returns the variance of a scalar, $\mathrm{Cov}(.)$ returns the covariance matrix of its vector argument, $\mathop{\mathbb{E}}[.]$ is the expectation operator, and $\mathrm{Tr}\{.\}$ is the matrix trace operator. The symbol $\mathcal{N}\left(\mu,\sigma\right)$ represents the normal (Gaussian) distribution with mean $\mu$ and standard deviation $\sigma$.

\section{Related Work}
We divide the related existing work into the following categories: 
    \begin{itemize}
        \item self-localization by a static blind node (single/multiple): self-localization by a single or multiple static blind nodes in a sensor network with computation load at the blind node itself;       
        \item target localization (single/multiple): localization of a single or multiple mobile/static blind nodes by a group of nodes with processing load either on a central facility or distributed among the group of nodes interested in the target localization; 
        \item joint multiple blind and anchor node localization: problem of refining the anchor position information in addition to the blind node localization with distributed or central processing; 
        \item self-localization by a mobile node: self-localization by a mobile node using information from other mobile nodes within its radio communication range and computation load only at the blind node.
    \end{itemize}

The first category \enquote{self-localization by a static blind node} relates to the static node localization in sensor networks such as those used in agriculture and environmental monitoring \cite{cardell2004field, hart2006environmental}. Generally, such networks contain a small number of nodes with independent positioning sensors such as GPS receivers. The rest of the nodes use some cooperative localization techniques to estimate their positions. An iterative multilateration technique in which three or more nodes compute their positions and then act as anchors in the subsequent iterations is given in \cite{savvides2001dynamic}. In \cite{Shang2004}, a multidimensional scaling technique that uses node connectivity information for localization is proposed. It presents two methods, one builds a global map directly and the other builds local maps then stitches them together to build a global map. The authors in \cite{Doherty2001} present a convex position estimation technique in WSN based exclusively on connectivity-induced constraints processed in a centralized resource center. They model the known peer-to-peer communication constraints as a set of geometric constraints on the node position and solve the problem using SDP.

In \cite{Zhu2009}, a range-based positioning technique is proposed that has a computational complexity lower than the least-square method. Their algorithm is based on the linearized range measurement equations and implementation of a WLS criterion in a computationally efficient way. Similarly, there are other works \cite{Feng2010,Picard2010} that focus on providing low-complexity solutions. The work of \cite{zheng2009robust} implements a robust joint localization and time synchronization in WSN with bounded anchor position uncertainties. It implements a robust joint estimator based on minimizing the worst-case mean square error and the solution is obtained by solving a SDP problem. Similarly, there are other RSSI-based node localization methods such as \cite{whitehouse2007practical, patwari2003using, kumar2009distance}. However, none of these works consider the problem of localizing blind nodes when perturbations are present in both RSSI and anchor position information.

The second category \enquote{target localization} has received considerable attention due to its wider application domain including target tracking, habitat monitoring, and military tracking \cite{szewczyk2004habitat, blazevic2005location, swami2007wireless, puccinelli2005wireless}. In \cite{Biswas2004}, the authors proposed an SDP relaxation-based method for the localization in ad-hoc WSNs. The approach is to convert the non-convex quadratic distance constraints into linear constraints by introducing a relaxation to remove the quadratic term from the constraint. In contrast to our problem, it assumes the availability of accurate anchor position information. A closed form solution and corresponding CRLB for localizing a stationary source/emitter based on time difference of arrival (TDOA) measurements is presented in \cite{Ho2004}. The authors assume normally-distributed error in both anchor position information and TDOA measurements. In contrast, we consider RSSI measurements modeled using a log-normal shadowing path loss model. An SOCP-based approach for sensor network localization with anchor position uncertainty is given in \cite{Naddafzadeh-Shirazi2014}. It presents a robust localization approach using maximum-likelihood criteria under an unbounded uncertainty model for the anchor position error. A distributed multidimensional scaling approach for localization in WSNs is proposed in \cite{costa2006distributed}. It weights the range measurements based on their accuracy to account for the communication constraints in a sensor network. 

The focus of the third category \enquote{joint multiple blind and anchor node localization} is on refining the anchor position information in addition to localizing a blind node. In \cite{srirangarajan2007distributed}, the authors propose a two-step distributed sensor network localization approach using SCOP. In the first step, the blind nodes compute their locations using available anchor node position information. In the second step, anchor nodes refine their position estimates. Node localization for the underwater and underground networks is considered in \cite{lui2009semi}. The authors propose an SDP-based localization algorithm in the presence of anchor position error and uncertainties in the signal propagation speed affecting the time of arrival (TOA)-based distance measurement. They also derive the CRLB for the corresponding problem.

In \cite{angjelichinoski2014spear, angjelichinoski2015cramer, denkovski2016geometric}, the authors propose an RSSI-based solution for joint location estimation of a source and multiple anchor nodes. They also provide the theoretical bounds for their solution as well as an interpretation of the theoretical bounds. The nature of the perturbations considered in \cite{angjelichinoski2014spear, angjelichinoski2015cramer, denkovski2016geometric} is the same as what we consider in this paper. However, their algorithm requires multiple RSSI and anchor position measurements. Nonetheless, our work is focused on energy-efficient outdoor localization of mobile nodes where the GPS is the main source of anchor position information while being responsible for the majority of energy consumed. Therefore, the requirement of multiple measurements of a anchor position is not in line with our goal of energy/resource-efficient localization. In a nutshell, this category is focused on Joint blind node-anchor node localization using complex optimization techniques such as SOCP and SDP, generally considered unsuitable for energy- and resource-constrained scenarios.

The last category is \enquote{self-localization by a mobile node}. The example application cases for this category are proximity services, mobile phone users tracking, and animal tracking with standalone tracking devices \cite{sommer2016lab, zhao2016understanding, raun2016measuring}. This category has the following features:
\begin{itemize}
\item Unlike in target localization, the anchor nodes are not required to be in the communication range of each other.
\item Limited resources highlight the need to avoid complex algorithms in performing multi-source or joint localization.
\item The uncontrolled mobile nature of the nodes leads to temporary grouping behavior. The grouping interval sometimes may not be enough to receive or share the results of joint localization. 
\end{itemize}

In \cite{jurdak2010adaptive}, the authors proposed a localization technique using RSSI based distance from neighboring nodes. They proposed to augment GPS-based positioning with more energy-efficient location sensors to bound position estimate uncertainty while the GPS is off. They used RSSI based distance estimation from the nodes with better position estimate as a measure to reduce self-position uncertainty. In \cite{vukadinovic2012performance}, the authors explored the concept of collaborative GPS duty cycling using Wi-Fi ad-hoc connectivity while ensuring application specific error bounds. Whenever a node approaches the error bound, first it  requests a better position estimate from its neighborhood and wait for a response within a certain time. In the case of a positive response, the node will update its position estimate. Otherwise, it will go for its own GPS lock followed by a broadcast of new GPS position coordinates.

In \cite{jurdak2010adaptive, vukadinovic2012performance}, the authors present an algorithm for self-localization in resource-constrained environments but their main focus is to minimize the energy consumption. They neither consider the problem of perturbation in distance estimation nor in the neighborhood position information. In \cite{Taniuchi2015}, the authors focus on improving the Wi-Fi positioning in indoor environments by using the RSSI-based distances among neighbors. They calculate confidence scores as weights in deciding the positions of the neighbors. The confidence of the Wi-Fi-based position is a function of the standard deviation of the multiple Wi-Fi scans for the same point. A lower standard deviation means higher confidence on position and vice-versa. Similarly, the confidence score of the Bluetooth is assigned by RSSI modeling in different settings of the indoor environment. Lastly, game theory is used to determine the final position of the nodes. Similar to others, this work only considered the error in the RSSI measurements. A simple subspace-based algorithm for single mobile node positioning using TOA measurements from three or more anchors with exact position information is given in \cite{so2007generalized}. 

In \cite{Tarrio2011}, the authors propose two WLS-based algorithms, called hyperbolic and circular, to localize a node in the presence of log-normal perturbations in the RSSI-based distance measurements. The hyperbolic algorithm linearizes the problem and solves it using the WLS method. The circular algorithm  minimize the weighted approximation of the original non-linear sum-square-error cost function using the gradient-descent method. The circular algorithm performs better than the hyperbolic algorithm due to minimization of the original cost function by an iterative approach. The proposed algorithms match the low computational requirement of our applications of interest but do not consider perturbations in the anchor node positions. Our proposed algorithm is based on a similar approach of non-linear WLS type but considers the error/noise in the anchor positions as well.

In summary, the existing relevant works either assume the perturbations in both RSSI-based distance measurements and anchor positions to have normal distribution, which is not realistic given that the RSSI-based distance measurements follow log-normal distribution \cite{rappaport1996wireless}, or propose solutions that are based on complex optimization techniques such as SDP or SOCP. The applications of our interest fall into the category of \enquote{self-localization by a mobile node} where a resource-constrained mobile node has access to perturbed anchor node positions and RSSI measurements and is interested in localizing itself in a resource-efficient way. We consider the more realistic log-normal perturbation for the RSSI-based distance measurements and normal distribution for anchor position perturbations. To meet the resource-efficiency requirement, we find a weighted nonlinear least-squares solution for the considered location estimation problem by minimizing a weighted sum-square error cost function using the gradient-descent method.   

\section{Problem Statement}

We consider the problem of self-localization by a single node, referred to as the blind node, on a two-dimensional Cartesian plane. The blind node is interested in obtaining an estimate of its true position, denoted by $(x_b,y_b)$. There are $M\geq3$ nodes, referred to as the anchor nodes, arbitrarily distributed within the communication range of the blind node at locations $(x_i,y_i)$, $i=1,...,M$. The locations of the anchor nodes are known to the blind node only approximately as they are corrupted by random perturbations. The blind node reckons its distance from the anchor nodes using available RSSI measurements that are also subject to random perturbations. We denote the perturbed knowledge of the anchor node positions at the blind node by $(\tilde{x}_{i},\tilde{y}_{i})$, $i=1,...,M$, and the corresponding perturbed RSSI measurements by $\tilde{p_{i}}$, $i=1,...,M$, in the linear (mW) scale and by $\tilde{p}_{i\text{(dBm)}}$, $i=1,...,M$, in the logarithmic (dBm) scale.

We adopt the following common assumptions:

\textit{A1}: The available knowledge of the position of the $i$th anchor node on $x$ and $y$ axes are corrupted by independent additive zero-mean Gaussian perturbations with standard deviation $\sigma_{a_{i}}$. The perturbations in the knowledge of different anchor node positions are independent of each other and the values of $\sigma_{a_{i}} $ may not be the same for different anchor nodes. Therefore, we have
\begin{equation*}
\tilde{x}_i = x_{i} + {n}_{x_i}
\end{equation*} 
\begin{equation*}
\tilde{y}_i = y_{i} + {n}_{y_i}
\end{equation*}    
\begin{equation*}
{n}_{x_i}, {n}_{y_i} \sim \mathcal{N}(0,\sigma_{a_{i}}).
\end{equation*}    

 

\textit{A2}: The path-loss model for the radio signal propagation is the log-normal shadowing model. Therefore, the RSSI measurement of the signal transmitted from the $i$th anchor node and received at the blind node has a nominal value of $\bar{p}_{i\text{(dBm)}}$ in the logarithmic (dBm) domain. However, the actual measured value is a realization of the nominal value corrupted by a zero-mean Gaussian perturbation with standard deviation $\sigma_{p_{i}}$, i.e.,\begin{equation}\label{pidbm1}
\tilde{p}_{i\text{(dBm)}}=\bar{p}_{i\text{(dBm)}}+n_{p_i}
\end{equation}
\begin{equation*}
n_{p_i} \sim \mathcal{N}\left(0,\sigma_{p_{i}}\right).
\end{equation*}

According to the shadowing path-loss model, we have
\begin{equation}\label{pidbm2}
\bar{p}_{i\text{(dBm)}} = p_{0\text{(dBm)}} - 10 \eta\ln{\frac{d_i}{d_0}}
\end{equation}
where
\begin{equation*}
d_i=\sqrt{\left(x_i-x_b\right)^2+\left(y_i-y_b\right)^2}
\end{equation*}
is the distance between the blind node and the $i$th anchor node. In addition, $d_0$, $p_{0\text{(dBm)}}$, and $\eta$, are the reference distance, the received power at the reference distance, and the path-loss exponent, respectively. Therefore, given the perturbed value $\tilde{p}_{i\text{(dBm)}}$, the RSSI-induced estimate for the distance between the blind node and the $i$th anchor node, denoted by $\tilde{d}_{i}$, is given by 
\begin{equation*}
\tilde{d}_i = d_0 10^{\cfrac{\tilde{p}_{i\text{(dBm)}} - p_{0\text{(dBm)}}}{10\eta}}.
\end{equation*}

Furthermore, we assume that the blind node and the anchor nodes have limited computational and energy resources. Hence, at any particular instance of localization, only one RSSI measurement and position estimate from each anchor node is available to the blind node.

\section{Proposed Algorithm}

One can estimate the position of the blind node by minimizing the following sum-square-error cost function

\begin{equation*}
c(x,y)  = \sum_{i= 1}^{M} {e_i}^2(x,y)
\end{equation*}
where
\begin{equation*}
e_i(x,y) = \delta_i - d_i
\end{equation*}
and
\begin{equation*}
\delta_i=\sqrt{(x_i - x)^2 + (y_i -y)^2}.
\end{equation*}

Given an initial estimate, the blind node's position can be estimated using a gradient-descent method as follows:

\begin{equation}\label{SGD}
\begin{split}
&\hat{x}_b^{(k+1)} = \hat{x}_b^{(k)} - \alpha{\frac{\partial c}{\partial x}}\Bigr|_{x = \hat{x}_b^{(k)}}\\
&\hat{y}_b^{(k+1)}= \hat{y}_b^{(k)} - \alpha{\frac{\partial c}{\partial y}}\Bigr|_{y = \hat{y}_b^{(k)}}
\end{split}
\end{equation}
where $\alpha>0$ is the step-size parameter and $\hat{x}_b^{(k)}$ and $\hat{y}_b^{(k)}$ are the estimate of $x_b$ and $y_b$ at iteration $k$, respectively.

However, we do not have access to the unperturbed values $x_i$, $y_i$, and $d_i$. Hence, we replace them with their corresponding perturbed observations  $\tilde{x}_i$, $\tilde{y}_i$, and $\tilde{d}_i$ and approximate the $i$th error term with 
\begin{equation*}
\bar{e}_i(x,y) = \tilde{\delta}_i - \tilde{d}_i
\end{equation*}
where
\begin{equation*}
\tilde{\delta}_i=\sqrt{(\tilde{x}_i - x)^2 + (\tilde{y}_i -y)^2}
\end{equation*}
In addition, to factor in the difference in the scale and statistical properties of the values associated with different anchor nodes, we weight each error term with the inverse of its standard deviation. Therefore, we minimize the following WLS cost function
\begin{equation}\label{epsilon_hat}
\bar{c}(x,y) = \sum_{i=1}^M\frac{\bar{e_i}^2}{\mathrm{Var}(\bar{e}_i)}.
\end{equation}

Since the perturbations of the anchor positions and RSSI-induced distances are independent of among each other, each variance term $\mathrm{Var}(\bar{e_i})$ can be calculated as 
\begin{equation}\label{Var(e_hat)}
\mathrm{Var}(\bar{e_i})=\mathrm{Var}(\tilde{\delta}_i) + \mathrm{Var}(\tilde{d_i}).
\end{equation}

To calculate the first term on the right-hand side of (\ref{Var(e_hat)}), we note that $\tilde{\delta}_i$ is the Euclidean distance of the points $(x,y)$ and $(\tilde{x}_i,\tilde{y}_i)$ where $x$ and $y$ are deterministic variables and $\tilde{x}_i$ and $\tilde{y}_i$ are independent stochastic variables that have Gaussian distributions with means $x_i$ and $y_i$, respectively, and the same variance $\sigma_{a_i}^2$. Therefore, $\tilde{\delta}_i$ has a Rice distribution \cite{Rice1945} with the variance
\begin{equation}\label{Var(w)-2}
    \mathrm{Var}(\tilde{\delta}_i)=\delta_i^2+2\sigma_{a_i}^2-\frac{\pi\sigma_{a_i}^2}{2}L_{1/2}^2\left(-\frac{\delta_i^2}{2\sigma_{a_i}^2}\right)
\end{equation}
where
\begin{equation*}
\delta_i=\sqrt{(x_i-x)^2 +(y_i-y)^2}
\end{equation*}
and $L_{1/2}(.)$ is a Laguerre polynomial expressed as 

\begin{equation*}
    L_{1/2}(z) =\exp\left(z/2\right)\left [\left( 1-z\right)I_0\left(-\frac{z}{2}\right)-zI_1\left (-\frac{z}{2}\right)\right]
\end{equation*}
with $I_0(.)$ and $I_1(.)$ being the modified Bessel functions of the first kind with order zero and one, respectively.

Considering the assumption \textit{A2}, the second term on the right-hand side of \eqref{Var(e_hat)} is calculated as \cite{Tarrio2011}
\begin{equation}\label{Var(d_tilde)}
\mathrm{Var}({\tilde{d_i}})=d_i^2\left[\exp\left({2\sigma_{d_i}^2}\right)-\exp\left({\sigma_{d_i}^2}\right)\right]
\end{equation}
where
\begin{equation*}
\sigma_{d_i} = \frac{\ln10}{10\eta}\sigma_{p_{i}}.
\end{equation*}
 




We estimate the blind node position by minimizing the formulated cost function \eqref{epsilon_hat} in an iterative manner using the gradient-descent method. Since the unperturbed anchor positions $(x_i,y_i)$, $i=1,...,M$, are unknown, we replace $\delta_i$ in \eqref{Var(w)-2} with its approximate value of
\begin{equation*}
\bar{\delta}_i^{(k)}=\sqrt{\left(\tilde{x}_i-\hat{x}_b^{(k)}\right)^2 +\left(\tilde{y}_i-\hat{y}_b^{(k)}\right)^2}
\end{equation*}
where $\left(\hat{x}_b^{(k)},\hat{y}_b^{(k)}\right)$ is the most recent estimate of the blind node location. In addition, we replace the unknown values of $d_i$, $i=1,...,M$ in (\ref{Var(d_tilde)}) with their available approximate values $\tilde{d}_i$, $i=1,...,M$. Therefore, we calculate the gradients as
\begin{equation*}
    \frac{\partial{\bar{c}}}{\partial x}\Bigr|_{x=\hat{x}_b^{(k)}} = -2\sum_{i=1}^M\frac{\left(\bar{\delta}_i^{(k)}-\tilde{d}_i\right)\left(\tilde{x}_i-\hat{x}^{(k)}_b\right)}{w_i^{(k)}\bar{\delta}_i^{(k)}}
\end{equation*}
and
\begin{equation*}
    \frac{\partial{\bar{c}}}{\partial y}\Bigr|_{y=\hat{y}_b^{(k)}} = -2\sum_{i=1}^M\frac{\left(\bar{\delta}_i^{(k)}-\tilde{d}_i\right)\left(\tilde{y}_i-\hat{y}^{(k)}_b\right)}{w_i^{(k)}\bar{\delta}_i^{(k)}}
\end{equation*}
where the weight $w_i^{(k)}$ that is an estimate of the variance of the error due to the measurements related to the $i$th anchor node is calculated at the $k$th iteration as
\begin{equation*}
\begin{split}
    w_i^{(k)}&=\left(\bar{\delta}_i^{(k)}\right)^2+2\sigma_{a_i}^2-\frac{\pi\sigma_{a_i}^2}{2}L_{1/2}^2\left(-\frac{\left(\bar{\delta}_i^{(k)}\right)^2}{2\sigma_{a_i}^2}\right)\\
    &+\tilde{d}_i^2\left[\exp\left({2\sigma_{d_i}^2}\right)-\exp\left({\sigma_{d_i}^2}\right)\right].
\end{split}
\end{equation*}

\subsection{Computational Complexity}
At each iteration, the proposed algorithm requires $16M$ multiplications and $4M$ divisions as well as $M$ square-root and $4M$ exponentiation operations. The required Bessel function values can be obtained using look-up tables, which only have some extra storage requirement. Therefore, the proposed algorithm has a computational complexity of $\BigO{IM}$ where $I$ is the total number of iterations needed for convergence.

\section{Cramer-Rao Lower Bound}

Given an observation vector $\mathbf{x}$ with known distribution that is related to an unknown parameter vector $\bm{\theta}$, the Cramer-Rao lower bound (CRLB) sets a lower bound on the covariance of any unbiased non-Bayesian estimator of $\bm{\theta}$. The CRLB is the inverse of the Fisher information matrix (FIM), denoted by $\mathbf{F}(\bm{\theta})$. Therefore, we have
\begin{equation*}
    \mathrm{Cov}(\hat{\bm{\theta}}) \geq \mathbf{F}^{-1}(\bm{\theta}).
\end{equation*}

The FIM represents the information provided by the observation $\mathbf{x}$  about the unobserved parameter vector $\bm{\theta}$ and is calculated as
\begin{equation*}
    \mathbf{F}(\bm{\theta}) = -\mathop{\mathbb{E}}\left[\frac{\partial^2
    l(\bm{\theta}\mid\mathbf{x})}{\partial^2\bm{\theta}}\right]
\end{equation*}
where $l(\bm{\theta}\mid\mathbf{x})$ is the log-likelihood function of $\bm{\theta}$ given $\mathbf{x}$.

In our self-localization problem, the observation vector $\mathbf{x}$ contains the perturbed anchor positions $ (\tilde{x}_i, \tilde{y}_i)$, $i=1,...,M$, as well as the perturbed RSSI measurements $\tilde{p}_i$, $i=1,...,M$, i.e., it can be written as
\begin{equation*}
\mathbf{x} = \left[\tilde{p}_1,\tilde{x}_1,\tilde{y}_1,\tilde{p}_2,\tilde{x}_2,\tilde{y}_2,...,\tilde{p}_M,\tilde{x}_M,\tilde{y}_M\right]^T.   
\end{equation*}

Given \eqref{pidbm1} and \eqref{pidbm2}, the probability density function (pdf) of $\tilde{p}_i$ is expressed as \cite{patwari2003relative}

\begin{equation*}\label{pdf_P}
f_{\tilde{p}_i}\left(\tilde{p}_i \right) = \frac{10/ \ln10 }{\sqrt{2\pi\sigma_{p_{i}}^2} \tilde{p}_i} \exp\left[-\frac{b_i}{8} \left(\ln\frac{d_i^2}{\tilde{d}_i^2}\right)^2 \right]    
\end{equation*}
 where
 \begin{equation*}
 b_i=\left(\frac{10 \eta}{\sigma_{p_{i}}\ln10} \right)^2.
 \end{equation*}
In view of the assumption \textit{A1}, the pdfs of the perturbed anchor positions $(\tilde{x}_i,\tilde{y}_i)$, $i=1,...,M$, are written as
\begin{equation*}
f_{\tilde{x}_i}(\tilde{x}_i) = \frac{1}{\sqrt{2\pi\sigma_{a_i}^2}} \exp\left[-\frac{(\tilde{x}_i -x_i)^2}{2\sigma_{a_i}^2}\right]
\end{equation*}
\begin{equation*}
f_{\tilde{y}_i}(\tilde{y}_i) = \frac{1}{\sqrt{2\pi\sigma_{a_i}^2}} \exp\left[-\frac{(\tilde{y}_i -y_i)^2}{2\sigma_{a_i}^2}\right].
\end{equation*}

Our unknown parameters of interest are the position coordinates of the blind node $(x_b,y_b)$. However, since the RSSI measurements are functions of the unknown unperturbed anchor positions, $(x_i,y_i)$, $i=1,...,M$, we include the unperturbed anchor positions as the nuisance parameters. Hence, our unknown parameter vector $\bm{\theta}$ is
\begin{equation*}
\bm{\theta} = \left[x_b,y_b,x_1,y_1,x_2,y_2,...,x_M,y_M\right]^T.
\end{equation*}

The perturbed anchor node positions $(\tilde{x}_i,\tilde{y}_i)$, $i=1,...,M$, are statistically independent of each other as well as the RSSI measurements. Therefore, we write the log-likelihood function as
\begin{equation*}\label{l(theta)}
\begin{split}
l \left(\bm{\theta}\mid\mathbf{x}\right)
&= \sum_{i=1}^M \ln f_{\tilde{p}_i}\left(\tilde{p}_i\mid\bm{\theta}\right) 
+\sum_{i=1}^M \ln f_{\tilde{x}_i}\left( \tilde{x}_i \mid\bm{\theta}\right)\\
&+ \sum_{i=1}^M \ln f_{\tilde{y}_i}\left( \tilde{y}_i\mid\bm{\theta}\right)\\
&= \sum_{i=1}^M \ln f_{\tilde{p}_i}\left(\tilde{p}_i\mid x_b,y_b,x_i,y_i\right)\\
& +\sum_{i=1}^M \ln f_{\tilde{x}_i}\left( \tilde{x}_i \mid x_i\right) + \sum_{i=1}^M \ln f_{\tilde{y}_i}\left( \tilde{y}_i\mid y_i\right)
\end{split}
\end{equation*}
and derive its second-order partial derivatives with respective to the entries of $\bm{\theta}$, required for the calculation of the FIM, in the Appendix. Using \cite{patwari2003relative}
\begin{equation*}
\mathbb{E}\left[\ln\left(\frac{d_i^2}{\tilde{d}^2_i}\right)\right] = 0
\end{equation*}
together with the results in the Appendix, we can express the FIM as
\begin{equation*}
\mathbf{F}(\bm{\theta}) = \begin{bmatrix}
\mathbf{F}_{11}&\mathbf{F}_{12}\\
\mathbf{F}_{12}^T&\mathbf{F}_{22}
\end{bmatrix}
\end{equation*}
where

\begin{equation*}
\mathbf{F}_{11}=\sum_{i=1}^M\frac{b_i}{d_i^4}\mathbf{Q}_i,
\end{equation*}

\begin{equation*}
\mathbf{F}_{12}=-\begin{bmatrix}
\frac{b_1}{d_1^4}\mathbf{Q}_1,\cdots,\frac{b_M}{d_M^4}\mathbf{Q}_M
\end{bmatrix},
\end{equation*}
\begin{equation*}
\mathbf{F}_{22}=\mathrm{blockdiag}\left\{\frac{b_1}{d_1^4}\mathbf{Q}_1+\frac{1}{\sigma_{a_1}^2}\mathbf{I}_2,\cdots,\frac{b_M}{d_M^4}\mathbf{Q}_M+\frac{1}{\sigma_{a_M}^2}\mathbf{I}_2\right\},
\end{equation*}

\begin{equation*}
\mathbf{Q}_i=\begin{bmatrix}
(x_i-x_b)^2&(x_i-x_b)(y_i-y_b)\\(x_i-x_b)(y_i-y_b)&(y_i-y_b)^2
\end{bmatrix},
\end{equation*}
and $\mathbf{I}_2$ is the $2\times2$ identity matrix. Consequently, we obtain a lower bound on the localization root-mean-square error (RMSE) of any unbiased estimator of the blind node location as below
\begin{equation*}
\resizebox{\linewidth}{!}{%
$\sqrt{\mathbb{E}\left[\left(\hat{x}_b-x_b\right)^2+\left(\hat{y}_b-y_b\right)^2\right]}\geq\sqrt{\mathrm{Tr}\left\{\left(\mathbf{F}_{11}-\mathbf{F}_{12}\mathbf{F}_{22}^{-1}\mathbf{F}_{12}^T\right)^{-1}\right\}}$%
}.
\end{equation*}

\section{Experimental Setup}
\begin{figure*}
    \centering
    \includegraphics[width=\textwidth]{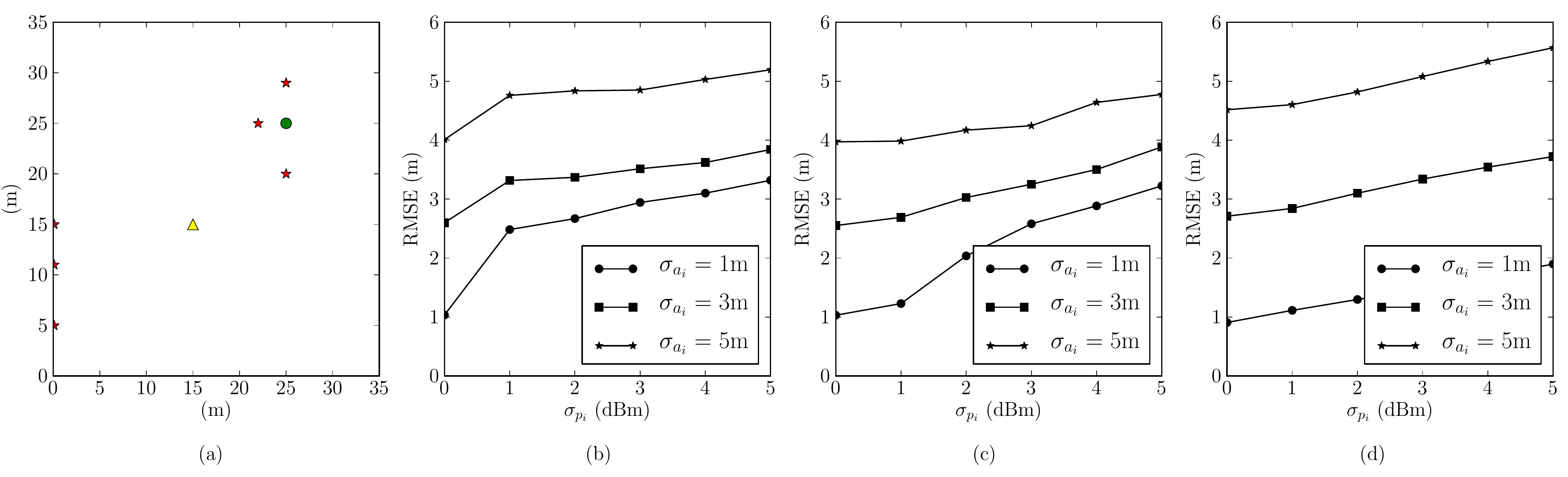}  
    \caption{Homogeneous GPS error, fixed nodes, semi-circle anchor node arrangement; (a) the network topology: the stars are the anchor nodes, the circle is the true position of the blind node, the triangle is the initial estimate for the blind node position; (b) the RMSE of the existing algorithm  for different values of $\sigma_{p_i}$ and $\sigma_{a_i}$; (c) the RMSE of the proposed algorithm  for different values of $\sigma_{p_i}$ and $\sigma_{a_i}$; (d) the corresponding CRLB values.}
    \label{fig:SHMS1}
    \vspace{5mm}
    
    \includegraphics[width=\textwidth]{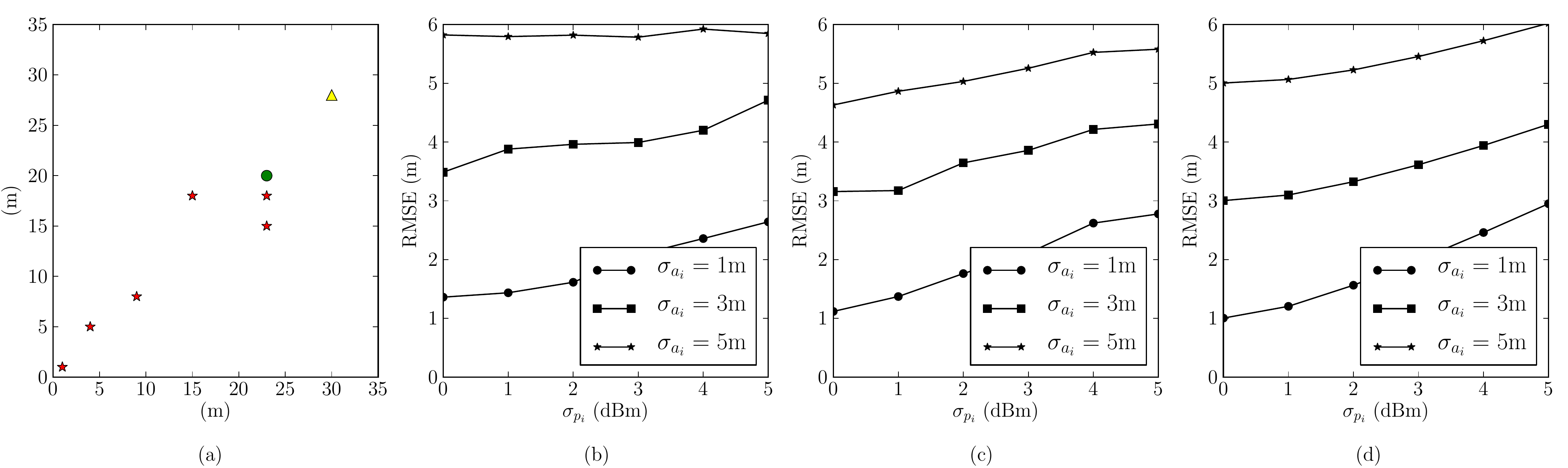}  
    \caption{Homogeneous GPS error, fixed nodes, semi-linear anchor node arrangement; (a) the network topology: the stars are the anchor nodes, the circle is the true position of the blind node, the triangle is the initial estimate for the blind node position; (b) the RMSE of the existing algorithm  for different values of $\sigma_{p_i}$ and $\sigma_{a_i}$; (c) the RMSE of the proposed algorithm  for different values of $\sigma_{p_i}$ and $\sigma_{a_i}$; (d) the corresponding CRLB values.}
    \label{fig:SHMS2}
\end{figure*}    

We consider the scenario of self-localization by a mobile blind node in the presence of six anchor nodes. All the nodes are assumed to be battery-powered and equipped with a processing unit and a low-power radio interface. Only anchor nodes are assumed to have a GPS unit for positioning. The perturbations on the position information of all anchor nodes are assumed to have Gaussian distribution with zero mean and known standard deviations. The radio propagation environment is assumed to be reasonably re-presentable by a log-normal shadowing path loss model with known parameters.

We evaluate the performance of the proposed algorithm in a $35\text{m}\times35\text{m}$ region. We consider both homogeneous and heterogeneous GPS noise scenarios in the region. The homogeneous scenario refers to the case where the standard deviation of the GPS noise is the same all over the region. In the heterogeneous scenario, the standard deviation of the GPS noise varies within the region. In practice, if the GPS operational context, e.g., surroundings, hardware, etc., is the same for all the nodes, we have a homogeneous scenario; otherwise, a heterogeneous one. As an example, the GPS localization performance is a function of the time during which the GPS receiver is active \cite{jurdak2010adaptive}. Therefore, variations in the GPS activity time may lead to a heterogeneous scenario. The heterogeneous scenario may also arise with cooperative or group/cluster-based energy-efficient localization schemes where the GPS activation time is decided based on the available energy budget of the individual nodes. In our experiments, the standard deviations of the perturbations on the anchor node position information, $\sigma_{a_i}$, $i=1,...,M$, are the same for all the anchor nodes when the GPS noise scenario is homogeneous. However, in the heterogeneous scenario, these standard deviations may differ depending on the position of the anchor nodes within the region.

The performance of the RSSI-based localization highly depends on the network geometry \cite{nguyen2016optimal,denkovski2016geometric}. Keeping this in mind, we further divide our experiments into two categories of \enquote{fixed} and \enquote{region-based random node placement}. In the fixed category, we evaluate the performance of the proposed algorithm with arbitrarily selected anchor and blind node positions, which we call  semi-linear, semi-circle, corner anchor, arbitrary node placements. We also consider several values for the standard deviation of the GPS noise, i.e., 1m, 3m, and 5m. The RSSI measurement errors range from 0dBm to 5dBm in all the experiments. To avoid any bias due to blind node position initialization, we select an arbitrary point in the region as the initial estimate for the blind node. The initialization point is kept considerably far away from the true position of the blind node while being fixed in all trials of each experiment.  

The region-based random node placement category is designed to emulate the geometries encountered by mobile nodes. In this category, we assign a region for the random placement of the anchor nodes. Similarly, we assign a region for the true position of the blind node as well as its initial position estimate. To further neutralize the impact of the network geometry on the localization performance, we consider two different anchor node placement arrangements, which we call corner and semi-circle, as well as varying the GPS noise standard-deviation within the range of $1-10$m. 

The values of the path-loss model parameters used in our experiments are $d_0=1$m, $p_{0(\text{dBm})}=-33.44$ and $\eta =3.567$. These values are based on the results reported in \cite{ahmad2015experiments}. We tune the step-size parameter of our gradient-descent algorithm to obtain convergence within $300$ iterations. We compare the performance of the proposed algorithm with that of the so-called weighted circular algorithm proposed in \cite{Tarrio2011}. This algorithm iteratively produces a WLS solution based on the assumption that only the RSSI measurements are corrupted by noise/error and the anchor node positions are exactly known at the blind node. Therefore, the proposed algorithm can be viewed as an improvement over the algorithm of \cite{Tarrio2011} taking into account the effects of perturbation in anchor node positions. We compare our algorithm only with this algorithm since, to the best of our knowledge, it is the only relevant algorithm in the literature that has a computational complexity comparable to that of the proposed algorithm. We use the same step-sizes and maximum number of iterations in the implementations of both algorithms. 

\section{Simulation Results}

\begin{figure*}
    \centering   
    \includegraphics[width=\textwidth]{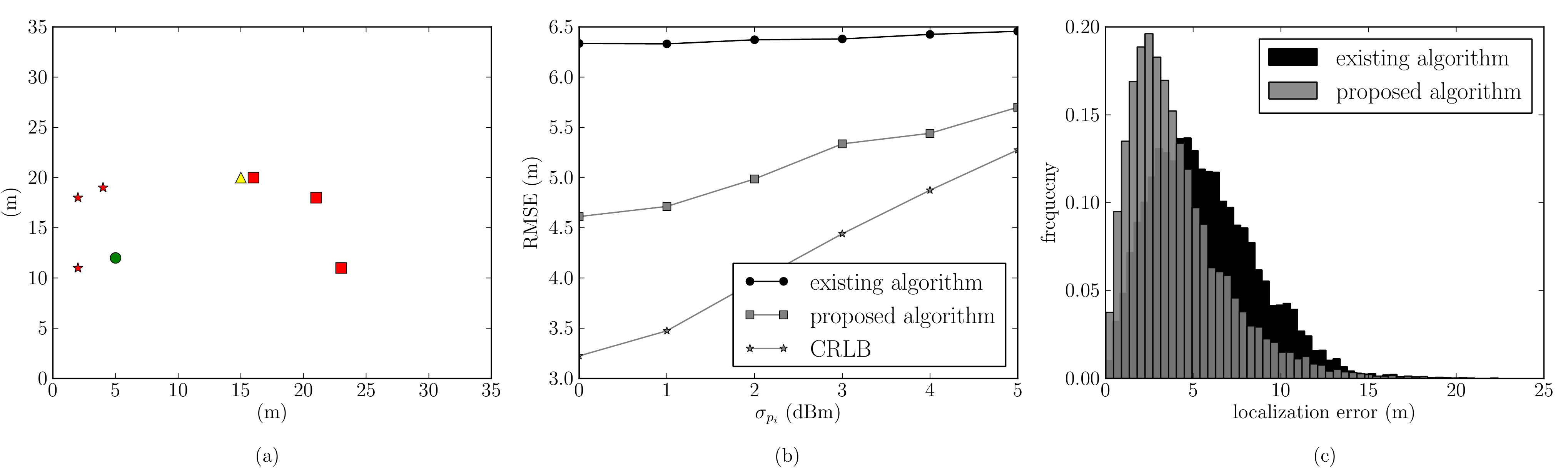}  
    \caption{Heterogeneous GPS error, fixed nodes, and arbitrary node placement; (a) the network topology: the stars are the anchor nodes with $\sigma_{a_i}=6$m, the squares are the anchor nodes with $\sigma_{a_i}=3$m, the circle is the true position of the blind node, the triangle is the initial estimate for the blind node position; (b) the RMSE of the proposed and existing algorithms as well as the corresponding CRLB for different values of $\sigma_{p_i}$; (c) the localization error histograms of the proposed and existing algorithms when $\sigma_{p_i} = 2$dBm.}
    \label{fig:SHTS1}
    \vspace{5mm}  
   
    \includegraphics[width=\textwidth]{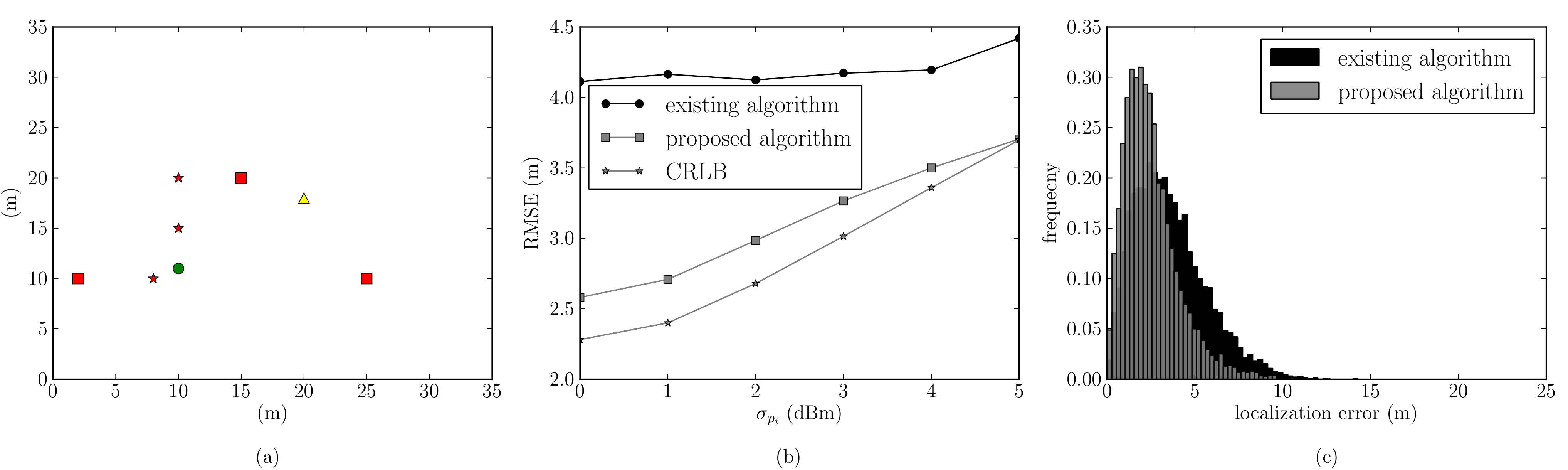}  
    \caption{Heterogeneous GPS error, fixed nodes, and arbitrary node placement; (a) the network topology: the stars are the anchor nodes with $\sigma_{a_i}=4$m, the squares are the anchor nodes with $\sigma_{a_i} = 2$m, the circle is the true position of the blind node, the triangle is the initial estimate for the blind node position; (b) the RMSE of the proposed and existing algorithms as well as the corresponding CRLB for different values of $\sigma_{p_i}$; (c) the localization error histograms of the proposed and existing algorithms when $\sigma_{p_i} = 2$dBm.}
    \label{fig:SHTS2}
    \vspace{5mm}
    
    \includegraphics[width=\textwidth]{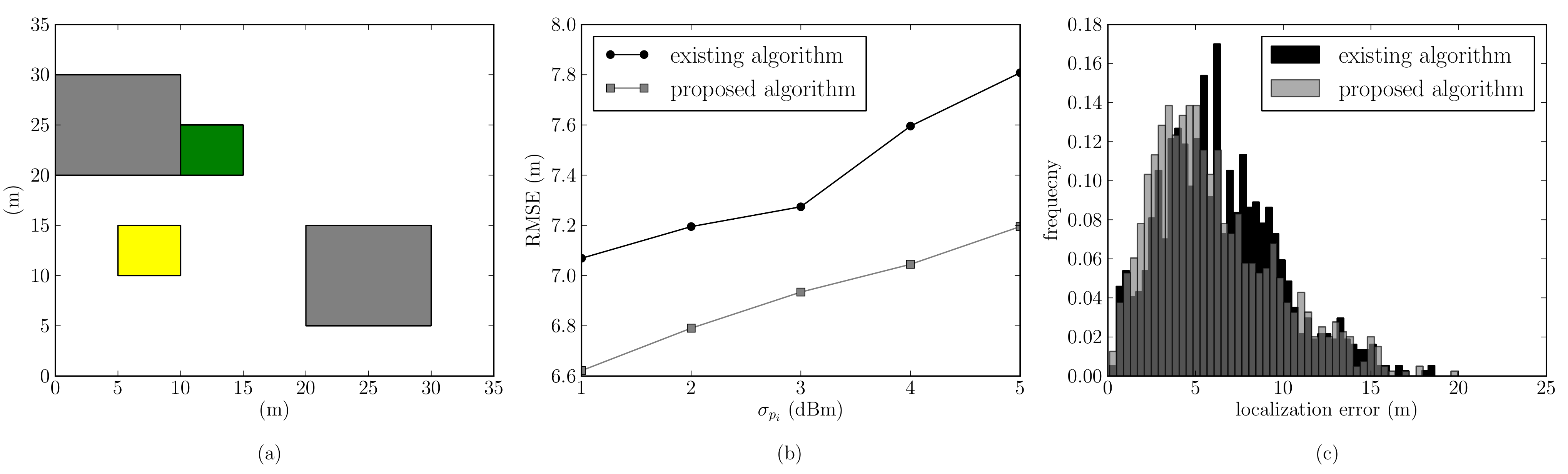}  
    \caption{Homogeneous GPS error, region-based random node placement, corner anchor node arrangement; (a) the network topology: the grey rectangles contain the anchor nodes with $\sigma_{a_i} = 5$m, the green rectangle contains the true position of the blind node, the yellow rectangle contains the initial estimate for the blind node position; (b) the RMSE of the proposed and existing algorithms; (c) the localization error histograms of the proposed and existing algorithms when $\sigma_{p_i} = 3$dBm.}
    \label{fig:MHMS1}
\end{figure*} 

\begin{figure*}  
    \centering    
    \includegraphics[width=\textwidth]{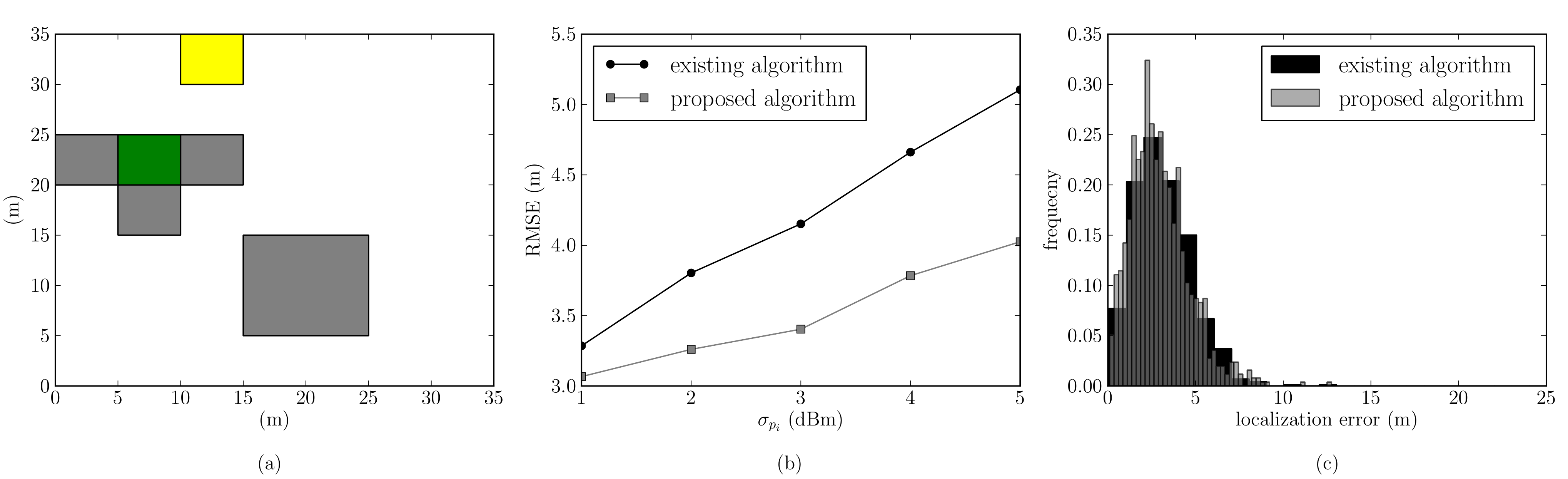}  
    \caption{Homogeneous GPS error, region-based random node placement, semi-circle anchor node arrangement; (a) network topology: the grey rectangles contain the anchor nodes with $\sigma_{a_i} = 3$m , the green rectangle contains the true position of the blind node, the yellow rectangle contains the initial estimate for the blind node position; (b) the RMSE of the proposed and existing algorithms; (c) the localization error histograms of the proposed and existing algorithms when $\sigma_p = 3$dBm.}
    \label{fig:MHMS2}
    \vspace{5mm}
    
    \includegraphics[width=\textwidth]{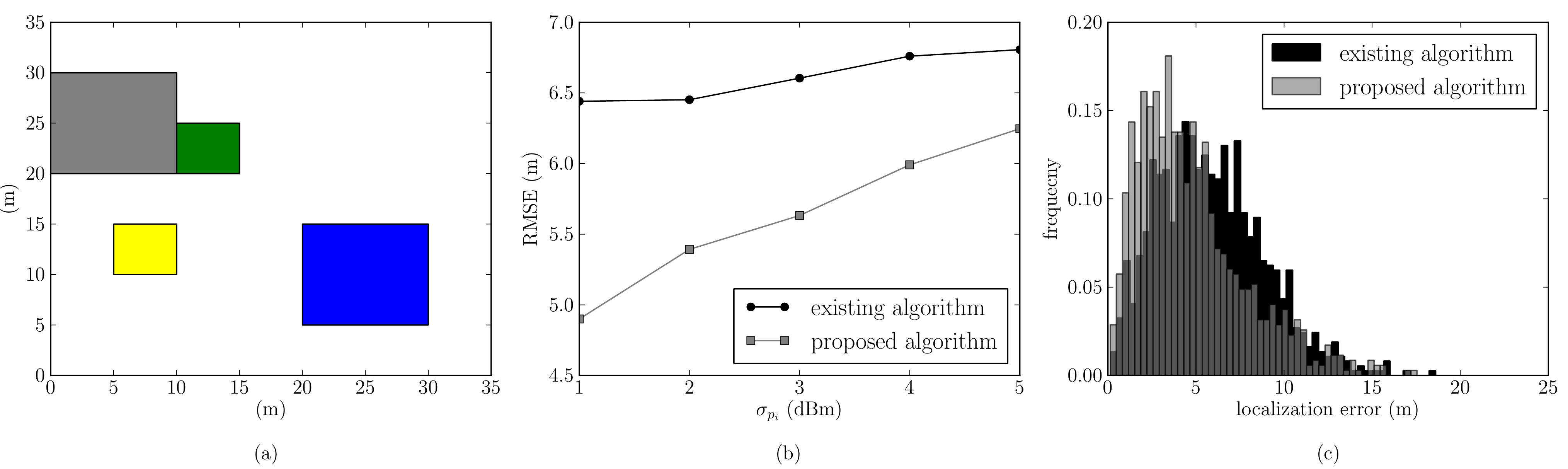}  
    \caption{Heterogeneous GPS error, region-based random node placement, corner anchor node arrangement; (a) network topology: the grey rectangle contains the anchor nodes with $\sigma_{a_i} = 5$m, the blue rectangle contains the anchor nodes with  $\sigma_{a_i} = 1$m, the green rectangle contains the true position of the blind node, the yellow rectangle contains the initial estimate for the blind node position; (b) the RMSE of the proposed and existing algorithms; (c) the localization error histograms of the proposed and existing algorithms when $\sigma_p = 3$dBm.}
    \label{fig:MHTS1}
    \vspace{5mm}
    
    \includegraphics[width=\textwidth]{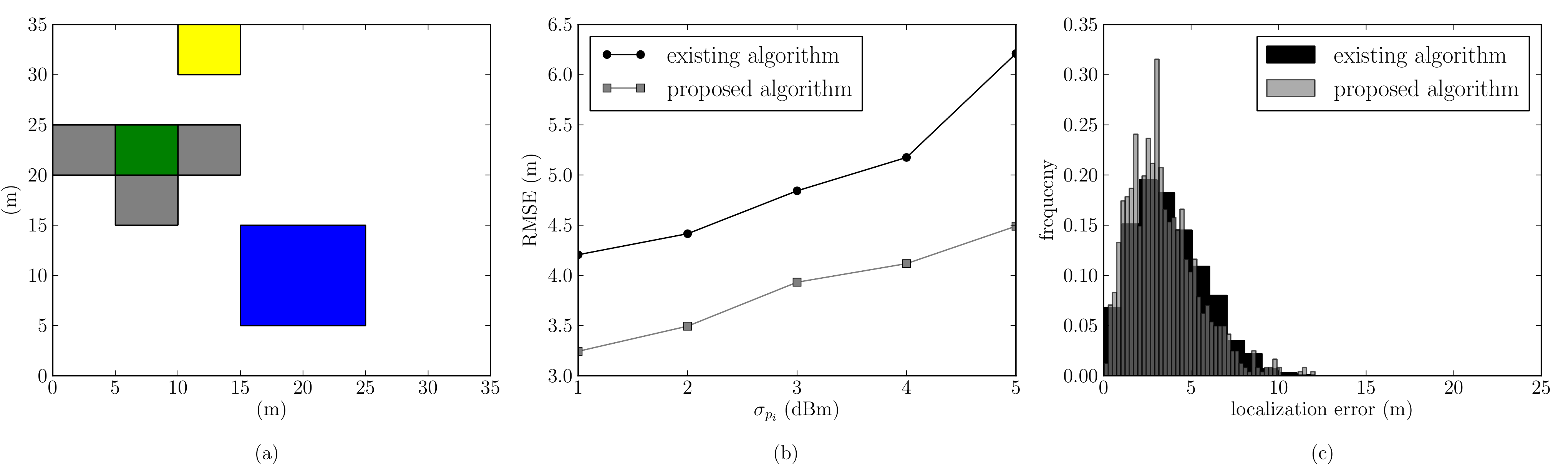}  
    \caption{Heterogeneous GPS error, region-based random node placement, semi-circle anchor node arrangement; (a) network topology: the grey rectangles contain the anchor nodes with $\sigma_{a_i} = 4$m, the blue rectangle contains the anchor nodes with $\sigma_{a_i} = 1$m, green rectangle contains the true position of the blind node, the yellow rectangle contains the initial estimate for the blind node position; (b) the RMSE of the proposed and existing algorithms; (c) the localization error histograms of the proposed and existing algorithms when $\sigma_p = 3$dBm.}
    \label{fig:MHTS2}
\end{figure*} 

We use the RMSE as the performance measure and calculate it by averaging over $1000$ independent trials for each experiment. We also present the corresponding values of the theoretical lower bound on the RMSE, when appropriate, although the estimates produced by the proposed algorithm may be biased in some scenarios.

\subsection{Fixed Node Placement} 

Fig. \ref{fig:SHMS1} presents the results for a homogeneous semi-circle arrangement with various values of $\sigma_{a_i}$ and $\sigma_{p_i}$. For $\sigma_{a_i}=5$m and $\sigma_{p_i}$ in the range of $1$dBm to $3$dBm, the proposed algorithm performs around $20\%$ better than the algorithm of \cite{Tarrio2011}. However, for other combinations of $\sigma_{a_i}$ and $\sigma_{p_i}$, both algorithms perform nearly similar except for some benefits of the proposed algorithm when $\sigma_{a_i} = 1$m and $\sigma_{p}=1, 2$dBm. Fig. \ref{fig:SHMS2} shows the results for a semi-linear arrangement with a noise environment similar to the above semi-circle arrangement. The proposed algorithm performs around $15\%$ better than the algorithm of \cite{Tarrio2011} for the combination of the higher range of $\sigma_{a_i}$ and the middle range of $\sigma_{p_i}$. 

Fig. \ref{fig:SHTS1} shows the results for a heterogeneous arbitrary selected node placement. In this experiment, the anchor nodes in the area with high GPS noise have $\sigma_{a_i}$ = $6$m and the other ones have $\sigma_{a_i}$ = $3$m. Overall, the proposed algorithm reduces the RMSE by $15\%$ to $30\%$. The proposed algorithm reduces the gap between the algorithm of \cite{Tarrio2011} and the CRLB by around $50\%$. 

The results for another arbitrarily selected node placement with variation in $\sigma_{a_i}$ is given in Fig. \ref{fig:SHTS2}. There are three anchor nodes with $\sigma_{a_i}=4$m and three anchor nodes with $\sigma_{a_i}=2$m. In low-noise radio environments, the RMSE of the proposed algorithm is just over half of that of the algorithm of \cite{Tarrio2011}. The proposed algorithm performs relatively close to the CRLB compared with the algorithm of \cite{Tarrio2011}. The error histogram of the algorithm of \cite{Tarrio2011} is also wider than that of the proposed algorithm for $\sigma_{p_i}=2$dBm.

\subsection{Region-Based Random Node Placement} 

Fig. \ref{fig:MHMS1} presents the results for the corner anchor node placement with homogeneous GPS error where the anchor nodes are randomly placed in two pre-specified regions. The values of $\sigma_{a_i}$ are 5m in both regions and each region has three anchor nodes. The RSSI error $\sigma_{p_i}$ varies from $1$dBm to $5$dBm. The proposed scheme reduces the RMSE by around 0.5m for all the variations of RSSI error. Fig. \ref{fig:MHMS2} shows a semi-circle arrangement of the nodes with $\sigma_{a_i} = 3$m. In this arrangement, the proposed scheme improves the RMSE around $10$ to $20\%$. 

The results for a corner anchor node placement arrangement with $\sigma_{a_i}=1$m and $5$m, respectively, for the low and high GPS error areas are given in Fig. \ref{fig:MHTS1}. The proposed scheme reduces the RMSE by approximately $25\%$ in the low RSSI error areas. With a high RSSI error, e.g., $\sigma_{p_i}= 5$dBm, the proposed algorithm still shows an improvement of $7\%$. The error histogram of the proposed algorithm for $\sigma_{p_i}=3$dBm is also slightly compacter than that of the algorithm of \cite{Tarrio2011}. 

Fig. \ref{fig:MHTS2} shows the results for a heterogeneous semi-circle arrangement. The standard-deviation of the anchor position error $\sigma_{a_i}$ is $4$m in the poor GPS performance area and $1$m in the high performance area. The proposed scheme performs $20$ to $25\%$ better in term of the RMSE with different RSSI error values. 

In summary, the proposed algorithm significantly outperforms the algorithm of \cite{Tarrio2011} in the realistic scenarios of heterogeneous GPS error. However, its advantages in the homogeneous GPS error scenarios is less pronounced.

\section{Conclusion}
We proposed a localization algorithm in the presence of the perturbation in the RSSI-based distance measurements as well as anchor positions information. We also derived the CRLB for the given problem. We evaluated the performance of the proposed algorithm in comparison with a previously-proposed algorithm that only accounts for perturbations in the RSSI measurements considering several arbitrary arrangements of anchor nodes and the blind node. Our simulation results showed that the proposed algorithm can significantly reduce the localization RMSE with respect to the existing algorithm.


\section{Appendix}

The second-order partial derivatives of the log-likelihood function $l\left(\bm{\theta}\mid\mathbf{x}\right)$ with respect to the entries of its argument $\bm{\theta}$, which are required for the calculation of the FIM, are computed as in the following:

\begin{equation*}
\resizebox{\linewidth}{!}{%
$\frac{\partial^2{l\left(\bm{\theta}\mid\mathbf{x}\right)}}{\partial{x_b}^2} 
= \sum_{i=1}^M\Bigg\{-\frac{b_i(x_i -x_b)^2}{d_i^4}\\
 +b_i\ln\left(\frac{d_i^2}{\tilde{d}_i^2} \right) \left[\frac{(x_i-x_b)^2}{d_i^4}-\frac{1}{2d_i^2}\right]\Bigg\}$%
}
\end{equation*}

\begin{equation*}
\resizebox{\linewidth}{!}{%
$\frac{\partial^2{l\left(\bm{\theta}\mid\mathbf{x}\right)}}{\partial{y_b}^2} 
= \sum_{i=1}^M\Bigg\{-\frac{b_i(y_i -y_b)^2}{d_i^4}\\
 +b_i\ln\left(\frac{d_i^2}{\tilde{d}_i^2} \right) \left[\frac{(y_i -y_b)^2}{d_i^4}-\frac{1}{2d_i^2} \right]\Bigg\}$%
}
\end{equation*}

\begin{equation*}
\begin{split}
\frac{\partial^2{l\left(\bm{\theta}\mid\mathbf{x}\right)}}{\partial x_b\partial y_b} = \sum_{i=1}^M\left\{\frac{b_i(x_i-x_b)(y_i- y_b)}{d_i^4}\left[\ln\left(\frac{d_i^2}{\tilde{d}_i^2} \right) - 1\right]\right\}
\end{split}
\end{equation*}

\begin{equation*}
\resizebox{\linewidth}{!}{%
$\frac{\partial^2{l\left(\bm{\theta}\mid\mathbf{x}\right)}}{\partial x_b\partial x_i}
= \frac{b_i(x_i -x_b)^2}{d_i^4}\\
 -b_i\ln\left(\frac{d_i^2}{\tilde{d}_i^2} \right) \left[\frac{(x_i -x_b)^2}{d_i^4} -\frac{1}{2d_i^2} \right]$%
}
\end{equation*}

\begin{equation*}
\frac{\partial^2{l\left(\bm{\theta}\mid\mathbf{x}\right)}}{\partial x_b \partial y_i}=\frac{b_i(x_i-x_b)(y_i-y_b)}{d_i^4}\left[1-\ln\left(\frac{d_i^2}{\tilde{d}_i^2}\right)\right]
\end{equation*}

\begin{equation*}
\frac{\partial^2{l\left(\bm{\theta}\mid\mathbf{x}\right)}}{\partial y_b \partial x_i}=\frac{b_i(x_i-x_b)(y_i-y_b)}{d_i^4}\left[1-\ln\left(\frac{d_i^2}{\tilde{d}_i^2}\right)\right]
\end{equation*}

\begin{equation*}
\resizebox{\linewidth}{!}{%
$\frac{\partial^2{l\left(\bm{\theta}\mid\mathbf{x}\right)}}{\partial y_b \partial y_i} 
= \frac{b_i(y_i-y_b)^2}{d_i^4} 
 -b_i\ln\left(\frac{d_i^2}{\tilde{d}_i^2} \right) \left[\frac{(y_i-y_b)^2}{d_i^4} -\frac{1}{2d_i^2} \right]$%
}
\end{equation*}

\begin{equation*}
\resizebox{\linewidth}{!}{%
$\frac{\partial^2{l\left(\bm{\theta}\mid\mathbf{x}\right)}}{\partial{x_i}\partial{x_j}}=\Bigg\{\begin{matrix} -\frac{b_i(x_i-x_b)^2}{d_i^4}-\frac{1}{\sigma_{a_i}^2}+b_i\ln\left(\frac{d_i^2}{\tilde{d}_i^2}\right)\left[\frac{(x_i-x_b)^2}{d_i^4}-\frac{1}{2d_i^2}\right]&i=j\\0&i\neq j
\end{matrix}$%
}
\end{equation*}

\begin{equation*}
\resizebox{\linewidth}{!}{%
$\frac{\partial^2{l\left(\bm{\theta}\mid\mathbf{x}\right)}}{\partial{y_i}\partial{y_j}}=\Bigg\{\begin{matrix}
-\frac{b_i(y_i-y_b)^2}{d_i^4}-\frac{1}{\sigma_{a_i}^2}+b_i\ln\left(\frac{d_i^2}{\tilde{d}_i^2}\right)\left[\frac{(y_i-y_b)^2}{d_i^4}-\frac{1}{2d_i^2}\right]&i=j\\0&i\neq j
\end{matrix}$%
}
\end{equation*}

\begin{equation*}
\resizebox{\linewidth}{!}{%
$\frac{\partial^2{l\left(\bm{\theta}\mid\mathbf{x}\right)}}{\partial x_i \partial y_j}=\Bigg\{\begin{matrix}
\frac{b_i}{d_i^4}(x_i-x_b)(y_i-y_b)\left[\ln\left(\frac{d_i^2}{\tilde{d}_i^2}\right)-1\right]&i=j\\0&i\neq j.
\end{matrix}$%
}
\end{equation*}

\bibliographystyle{IEEEtran}
\bibliography{References} 

\end{document}